\def\b{\beta}\def\c{\chi}
\def\f{\phi}\def\g{\gamma}
\def\l{\lambda}\def\m{\mu}\def\q{\psi}\def\t{\tau}

\def\L{\Lambda}
\def\O{\Omega}

\def\de{\partial}
\def\inf{\infty}\def\mo{{-1}}\def\ha{{1\over 2}}

\def\({\left(}\def\){\right)}\def\[{\left[}\def\]{\right]}

\def\const{{\rm const}}

\def\ep{\e_{\m\n}}

\def\fe{field equations }

\def\cco{cosmological constant }\def\em{electromagnetic }

\def\ads{anti-de Sitter }

\def\des{de Sitter }

\def\wrt{with respect to }

\def\section#1{\bigskip\noindent{\bf#1}\smallskip}

\def\PRL#1{Phys.\ Rev.\ Lett.\ {\bf#1}}
\def\PR#1{Phys.\ Rev.\ {\bf#1}}

\def\JHEP#1{JHEP\ {\bf#1}}
\def\RMP#1{Rev.\ Mod.\ Phys.\ {\bf#1}}
\def\arx#1{{\tt arXiv:#1}}

\def\ref#1{\medskip\everypar={\hangindent 2\parindent}#1}
\def\beginref{\begingroup
\bigskip
\centerline{\bf References}
\nobreak\noindent}
\def\endref{\par\endgroup}

\def\ep{e^{2\sqrt3\b\f}}\def\em{e^{2\sqrt3\f/\b}}
\def\ef{e^{2\sqrt3\f/\b}}
\magnification=1200

{\nopagenumbers
\line{}
\vskip60pt
\centerline{\bf An exactly solvable inflationary model}
\vskip60pt
\centerline{
{\bf S. Mignemi}$^*$\footnote{$^\ddagger$}{e-mail: smignemi@unica.it}
and {\bf N. Pintus}\footnote{$^\dagger$}{e-mail: nicola.pintus@hotmail.it}}
\vskip10pt
\centerline {Dipartimento di Matematica, Universit\`a di Cagliari}
\centerline{viale Merello 92, 09123 Cagliari, Italy}
\smallskip
\centerline{$^*$and INFN, Sezione di Cagliari}
\vskip80pt
\centerline{\bf Abstract}

\vskip10pt
{\noindent
We discuss a model of gravity coupled to a scalar field that admits exact cosmological
solutions displaying an inflationary behavior at early times and a power-law expansion
at late times.
}
\vskip80pt\
\vfil\eject}

\section{1. Introduction}
Inflationary models are nowadays the most successful attempt to describe the early
stages of development of the universe, and in particular are at the basis of the
current explanations of the formation of structures in the universe [1].
They are based on the hypothesis that a period of exponential expansion has taken
place shortly after the big bang.
Usually, their dynamics is derived from models of gravity minimally coupled to a
scalar field with suitable potential. The scalar field starts its evolution from a
value that does not minimize the potential and then rolls toward the minimum,
inducing the inflation of the scale factor of the universe.

In most cases,
these models can only be treated in an approximate way, in particular the so-called
slow-roll approximation is used to investigate their behavior. The approximation
breaks down when the field is close to the minimum of the potential, where the inflation
ends. The later evolution of the universe is assumed to obey the standard
Friedmann-Lema\^{\i}tre equations and is characterized by a power-law expansion.
It would therefore be interesting to find exact solutions that describe the
transition from an exponential expansion to a later Friedmann-Lema\^{\i}tre behavior
in a smooth way.

For this purpose, it may be useful to exploit the observation [2] that cosmological
solutions of models of gravity coupled to a scalar
can be obtained from domain-wall solutions of the same model with opposite
scalar potential, simply by analytic continuing to imaginary values the time
and radial coordinates. Recently, this procedure has been applied to a model with
a potential given by a sum of exponentials, which admits some exact solutions
displaying an accelerated expansion at late times [3].

In this paper we study another interesting application of this observation, leading to
an exact solution in which the scale factor expands exponentially at early times
and then passes to a power-law behavior.
This is based on a model of gravity minimally coupled to a scalar field $\f$,
introduced in [4] and defined by the action
$$I=\int\sqrt{-g}\[R-2(\de\f)^2-V(\f)\]d^4x,\eqno(1.1)$$
where the scalar potential
$$V(\f)=-{2\l^2\over3\g}\(\ep-\b^2\em\),\eqno(1.2)$$
depends on two parameters $\l$ and $\b$, with $\g=1-\b^2$.

It was shown in [4] that this model admits solitonic solutions of the form
$$ds^2=\hat R^{-{2\over1+3\b^2}}\(1+\m\hat R^{-{3\g\over1+3\b^2}}\)^{{2\b^2\over\g}}d\hat R^2
+\hat R^{2\over1+3\b^2}\(1+\m\hat R^{-{3\g\over1+3\b^2}}\)^{2\b^2\over3\g}(-d\hat T^2+
d\bar s_2^2),\eqno(1.3)$$
with $\m$ a free parameter, and $d\bar s_2^2$ the line element of flat 2-space.
This metric interpolates between \ads for $\hat R\to0$ and domain-wall behavior for
$\hat R\to\inf$.

Analytically continuing this solution for $\hat R\to iT$, $\hat T\to iR$, one can obtain
a cosmological solution for the action (1.1) with potential
$$V(\f)={2\l^2\over3\g}\(\ep-\b^2\em\),\eqno(1.4)$$
that reads
$$ds^2=-T^{-{2\over1+3\b^2}}\(1+\m T^{-{3\g\over1+3\b^2}}\)^{2\b^2\over\g}dT^2
+T^{2\over1+3\b^2}\(1+\m T^{-{3\g\over1+3\b^2}}\)^{2\b^2\over3\g}d\bar s_3^2,\eqno(1.5)$$
with  $d\bar s^2_3$ the line element of 3-dimensional flat space.
It is easy to verify that this solution behaves as a \des universe for $T\to0$ and as a Friedmann
universe with power-law expansion for $T\to\inf$.
It is then a promising candidate to describe an inflationary universe.

\section{2. General solutions}
The solution (1.5) is not the most general cosmological solution of the model (1.1) with potential
(1.4), and is therefore important to thoroughly investigate the system of equations derived from (1.1)
in order to see if the behavior of (1.5)
generic and to understand if it can describe a viable cosmological model.
As we shall see, the system is exactly integrable if a more suitable parametrization of the metric is
used rather than (1.5).
In this paper, we shall limit our study to solutions in absence of matter. We plan to extend our
investigation to the more general case of matter coupling in a further publication [5].

The potential (1.3) vanishes for $\f\to-\inf$, has a maximum for $\f=0$, where it takes
the value $V_0=2\l^2/3$, and goes to $-\inf$ for $\f\to\inf$.
Hence a solution with $\f=0$ exists, that coincides with that of pure gravity with cosmological
constant $\L=2\l^2/3$. This is of course the \des solutions with \cco $\L$.
It is interesting to notice that the potential (1.4) admits a duality for $\b\to1/\b$.
In the following we shall consider only the case $0<\b^2<1$. In the limit $\b\to0$, the potential
reduces to a cosmological constant, while for $\b^2=1$ it vanishes.
\bigskip

We are interested in the general isotropic and homogeneous cosmological solutions,
with flat spatial sections, that we parametrize as
$$ds^2=-e^{2a(\t)}d\t^2+e^{2b(\t)}d\bar s_3^2,\qquad\f=\f(\t),\eqno(2.1)$$
with $\t$ a time variable.

With this parametrization, the vacuum Einstein equations read
$$\eqalignno{&3\dot b^2=\dot\f^2+{V\over2}\,e^{2a},&(2.2)\cr
&2\ddot b+\dot b(3\dot b-2\dot a)=-\dot\f^2+{V\over2}\,e^{2a},&(2.3)}$$
where a dot denotes a derivative \wrt $\t$, while the scalar field obeys the
equation
$$\ddot\f+(3\dot b-\dot a)\dot\f=-{1\over4}{dV\over d\f}\,e^{2a}.\eqno(2.4)$$
In the gauge $b=a/3$ the previous equations simplify. In fact, with this choice,
they can be written as
$$\eqalignno{&{\dot a^2\over3}=\dot\f^2+\ha\,Ve^{2a},&(2.5)\cr
&\ddot a={3\over2}\,Ve^{2a},&(2.6)\cr
&\ddot\f=-{1\over4}{dV\over d\f}\,e^{2a}.&(2.7)}$$

It is now convenient to define new variables
$$\q=a+\sqrt3\b\f,\qquad\c=a+{\sqrt3\over\b}\f,\eqno(2.8)$$
such that
$$a={\q-\b^2\c\over\g},\qquad\f={\b(\c-\q)\over\sqrt3\g}.\eqno(2.9)$$

In terms of the new variables, the \fe take the very simple form
$$\ddot\q=\l^2e^{2\q},\qquad\ddot\c=\l^2e^{2\c},\eqno(2.10)$$
subject to the constraint
$$\dot\q^2-\b^2\dot\c^2=\l^2(e^{2\q}-\b^2e^{2\c}).\eqno(2.11)$$
The \fe are invariant under time reversal, $\t\to-\t$. Hence to each solution
corresponds a time-reversed one.

The equations (2.10) admit first integrals
$$\dot\q^2=\l^2e^{2\q}+Q_1,\qquad\dot\c^2=\l^2e^{2\c}+Q_2,\eqno(2.12)$$
with $Q_1$ and $Q_2$ integration constants, which according to (2.12) satisfy
$$Q_1=\b^2Q_2.\eqno(2.13)$$
The solutions of the system (2.10) depend on the sign of the integration constants.
If $Q_i=q_i^2>0$ ($i=1,2$),
$$\l^2e^{2\q}={q_1^2\over\sinh^2[q_1(\t-\t_1)]},\qquad\l^2e^{2\c}={q_2^2\over
\sinh^2[q_2(\t-\t_2)]},\eqno(2.14)$$
with $\t_1$, $\t_2$, integration constants and $q_1^2=\b^2q_2^2$.

If $Q_1=Q_2=0$,
$$\l^2e^{2\q}={1\over(\t-\t_1)^2},\qquad\l^2e^{2\c}={1\over(\t-\t_2)^2}.\eqno(2.15)$$
Finally, if $Q_i=-q_i^2<0$,
$$\l^2e^{2\q}={q_1^2\over\sin^2[q_1(\t-\t_1)]},\qquad\l^2e^{2\c}={q_2^2\over
\sin^2[q_2(\t-\t_2)]},\eqno(2.16)$$
again with $q_1^2=\b^2q_2^2$.
In the following we define $q=q_2={q_1/\b}$.

The solutions of the Friedman equations are therefore the following:

\noindent If $Q_i>0$,
$$e^{2a}={q^2\over\l^2}\ {\big|\,\sinh[q(\t-\t_2)]\,\big|^{2\b^2/\g}\over\big|\,
\b^\mo\sinh[\b q(\t-\t_1)]\,\big|^{2/\g}},\qquad
\ef=\left|{\,\sinh[\b q(\t-\t_1)]\over\b\sinh[q(\t-\t_2)]}\right|^{2/\g}.\eqno(2.17)$$

\noindent If $Q_i=0$,
$$e^{2a}={1\over\l^2}\ {\big|\t-\t_2\big|^{2\b^2/\g}\over\big|\t-\t_1\big|^{2/\g}},
\qquad\ef=\left|{\t-\t_1\over\t-\t_2}\right|^{2/\g}.\eqno(2.18)$$

\noindent If $Q_i<0$,
$$e^{2a}={q^2\over\l^2}\ {\big|\,\sin[q(\t-\t_2)]\,\big|^{2\b^2/\g}\over\big|\,
\b^\mo\sin[\b q(\t-\t_1)]\,\big|^{2/\g}},\qquad
\ef=\left|{\,\sin[\b q(\t-\t_1)]\over\b\sin[q(\t-\t_2)]}\right|^{2/\g}.\eqno(2.19)$$

Notice that (2.18) is nothing but the solution (1.5) in different coordinates.
In fact, they are related by the change of variables $\t=\const\times T^{-3\g/(1+3\b^2 )}$.

The actual value of the parameter $q$ is not important, it only sets the scale of the
time variable $\t$.

\section{3. Properties of the solutions}

To give a physical interpretation of the solutions, it is useful to define the cosmic time
$t$ such that $dt=\pm e^ad\t$. The possibility of choosing the plus or minus sign derives from
the invariance of the equations under time reversal. To each expanding solution therefore
corresponds an unphysical contracting solution, that we shall disregard.
In the new parametrization, the line element reads
$$ds^2=-dt^2+e^{2b(t)}d\O^2.\eqno(3.1)$$

Unfortunately, the solutions obtained in the previous section cannot be written in terms of
elementary functions of $t$, except when $Q_i=0$, $\t_1=\t_2$. In this case,
$t-t_0=\pm\log|\t-\t_1|$, with $t_0$ an arbitrary integration constant. Choosing the minus
sign in the previous expression, one obtains an expanding universe, with
$e^{2b}=e^{2(t-t_0)/3\l}$ and $\f=0$, namely a \des spacetime with vanishing scalar field. This is
of course the unstable solution corresponding to the scalar sitting at the top of the potential.
\bigskip
In the general case, the solutions have two qualitatively different branches if $\t_1=\t_2$, or
three if $\t_1\ne\t_2$. We are only interested in those branches where $t$ is a monotonic
function of $\t$ and the universe expands.
Studying their behavior for $\t\to\t_{1,2}$ and $\t\to\pm\inf$, we obtain the following
physically acceptable solutions, besides the one discussed above:

If $Q_i>0$ and $\t_1=\t_2$, for $t\to-\inf$, $e^{2b}\sim e^{2\l t/3}$ and $\f\sim\const$,
while for $t\to\inf$, $e^{2b}\sim t^{2/3}$ and $\f\sim\log t$.
This solution describes a universe with \des inflationary behavior for
$t\to-\inf$ that gradually turns to a power-law expansion for $t\to\inf$.

If $Q_i<0$ and $\t_1=\t_2$, for $t\to-\inf$, $e^{2b}\sim e^{2\l t/3}$ and $\f\sim\const$,
while for $t\to t_0$, $e^{2b}$ vanishes as $(t-t_0)^{2\b^2/3}$ and $\f\sim\log(t-t_0)$.
In this case another acceptable branch exists that behaves as $e^{2b}\sim(t-t_0)^{2\b^2/3}$ and
$\f\sim\log(t-t_0)$, both at $t=t_0$ and at a later finite time, describing an universe that initially
expands and then recontracts.

\bigskip
{\noindent If $\t_1\ne\t_2$, the solutions are more complicated and present two acceptable branches:}

For $Q_i=0$, one branch has behavior $e^{2b}\sim e^{2\l t/3}$ and $\f\sim\const$ for
$t\to-\inf$, while $e^{2b}\sim t^{2/3\b^2}$, $\f\sim\log t$ for $t\to\inf$.
The other branch behaves as $e^{2b}\sim(t-t_0)^{2\b^2/3}$ and $\f\sim\log(t-t_0)$\ for $t\to t_0$,
with $e^{2b}\sim t^{2/3\b^2}$, $\f\sim\log t\ $ for $t\to\inf$. In the first case the behavior is
qualitatively similar to the case in which $q\ne0$ and $\t_1=\t_2$.

For $Q_i>0$ in the first branch $e^{2b}\sim(t-t_0)^{2\b^2/3}$ and $\f\sim\log(t-t_0)$ for $t\to t_0$,
while $e^{2b}\sim t^{2/3}$, $\f\sim\log t$ for $t\to\inf$.
In the second branch, $e^{2b}\sim(t-t_0)^{2\b^2/3}$ and $\f\sim\log(t-t_0)$ for $t\to t_0$,  while
$e^{2b}\sim t^{2/3\b^2}$, $\f\sim\log t$ for $t\to\inf$.

For $Q_i<0$ the first branch has behavior $e^{2b}\sim(t-t_0)^{2\b^2/3}$ and $\f\sim\log(t-t_0)$ for
$t\to t_0$, with $e^{2b}\sim t^{2/3\b^2}$, $\f\sim\log t$ for $t\to\inf$. Another branch behaves as
$e^{2b}=(t-t_0)^{2\b^2/3}$ and $\f\sim\log(t-t_0)$, both at $t=t_0$ and at a later finite time.

\bigskip
We are mainly interested in those solutions that behave exponentially for $t\to-\inf$ and as a
power law for $t\to\inf$. These are obtained for $Q_i>0$, $\t_1=\t_2$, or $Q_i=0$, $\t_1\ne\t_2$.
They correspond to an initial configuration where the scalar field is in the unstable equilibrium
configuration at the top of the potential and then rolls down either to $-\inf$ or to $+\inf$.
The behavior of the two soutions is similar, but they differ for the exponent of the late power-law
expansion which is $2/3$ in the first case, and $2/3\b^2$ in the second.

The exponential expansion lasts until $\t=\t_f\sim1/q\b$, namely $t-t_0=t_f-t_0\sim1/\l$.
After this time the acceleration of the expansion becomes negative.
At such time the scale factor $e^{2b}$ is of order $\l^{-2/3}$. Denoting $t_i$ the time at which
the inflation starts, the scale factor therefore inflates by a factor $e^{-2\l(t_i-t_0)/3}$.
Choosing $t_i-t_0$ negative, one can then obtain the desired amount of inflation.

\section{4. Conclusion}
We have described an exactly solvable model of gravity minimally coupled to a scalar field,
which admits solutions presenting an initial inflationary period followed by a late power-law
expansion. These solutions correspond to a decay from an initial configuration when the scalar
field is in an unstable vacuum.
We have not included in the model ordinary matter, nor considered the stability of the inflationary
solutions with respect to the other solutions of the model.
We plan to discuss these topics in a future paper. We also intend to investigate more carefully if
the inflationary mechanism provided by this model is in accordance with the commonly accepted
mechanisms for the generation of inhomogeneities [1].
In this context, the existence of exact solutions might be helpful.
\bigbreak
\centerline{\bf Acknowledgments}
\smallskip
\noindent We wish to thank Mariano Cadoni for useful discussions.

\section{Appendix}
For completeness, in this section we briefly discuss the results obtained when the potential (1.2)
is used instead of (1.5). In this case the potential has a stable minimum for $\f=0$ and goes to
infinity for $\f\to\inf$. The properties of the solutions are therefore completely different.
However, the field equations (2.10), (2.11) are still valid, with the substitution $\l^2\to-\l^2$,
and their first integrals read
$$\dot\q^2=-\l^2e^{2\q}+q_1^2,\qquad\dot\c^2=-\l^2e^{2\c}+q_2^2,\eqno(A.1)$$
with positive integration constants satisfying $q^2_1=\b^2q^2_2$.
Integrating (A.1), one obtains
$$e^{2a}={q^2\over\l^2}\({\b\cosh^{\b^2}[q(\t-\t_2)]\over\cosh[\b q(\t-\t_1)]}\)^{2/\g},\qquad
\ef=\({\,\cosh[\b q(\t-\t_1)]\over\b\cosh[q(\t-\t_2)]}\)^{2/\g}.\eqno(A.2)$$
Contrary to the solutions of section 2, these solutions are regular everywhere, since the hyperbolic
cosine has no zeroes.

They represent universes starting with a big bang at $t=0$ and recontracting after a finite time.

\beginref
\ref [1] For a review, see
D.H. Lyth, A.R. Liddle, {\sl The primordial density perturbation}, Cambridge U.P. 2009;
B.A. Bassett, S. Tsujikawa and D. Wands, \RMP{78}, 537 (2006).
\ref [2] K. Skenderis and P. Townsend, \PRL{96}, 191301 (2006).
\ref [3] M. Cadoni and M. Ciulu, \arx{1311.4098}.
\ref [4] M. Cadoni, S. Mignemi and M. Serra, \PR{D85}, 086001 (2012);
M. Cadoni and S. Mignemi, \JHEP{1206}, 056 (2012).
\ref [5] M. Cadoni, S. Mignemi and N. Pintus, in preparation.
\endref
\end